\documentclass[sigconf]{acmart}
\usepackage{graphicx}
\usepackage{array}
\usepackage{subfig}

\AtBeginDocument{%
  \providecommand\BibTeX{{%
    \normalfont B\kern-0.5em{\scshape i\kern-0.25em b}\kern-0.8em\TeX}}}

\copyrightyear{2025}
\acmYear{2025}
\setcopyright{cc}
\setcctype{by}
\acmConference[CHI '25]{CHI Conference on Human Factors in Computing Systems}{April 26-May 1, 2025}{Yokohama, Japan}
\acmBooktitle{CHI Conference on Human Factors in Computing Systems (CHI '25), April 26-May 1, 2025, Yokohama, Japan}\acmDOI{10.1145/3706598.3713347}
\acmISBN{979-8-4007-1394-1/2025/04}

\begin{document}

\title[UX with LLM-powered CRS]{User Experience with LLM-powered Conversational Recommendation Systems: A Case of Music Recommendation}

\author{Sojeong Yun}
\affiliation{%
  \institution{Department of Industrial Design, KAIST}
  \city{Daejeon}
  \country{Republic of Korea}}
\email{thwjd3785@kaist.ac.kr}

\author{Youn-kyung Lim}
\affiliation{%
  \institution{Department of Industrial Design, KAIST}
  \city{Daejeon}
  \country{Republic of Korea}}
\email{younlim@kaist.ac.kr}

\renewcommand{\shortauthors}{Yun,S. and Lim,Y.}

\begin{abstract}\textcolor{black}{
    The advancement of large language models (LLMs) now allows users to actively interact with conversational recommendation systems (CRS) and build their own personalized recommendation services tailored to their unique needs and goals. This experience offers users a significantly higher level of controllability compared to traditional RS, enabling an entirely new dimension of recommendation experiences. Building on this context, this study explored the unique experiences that LLM-powered CRS can provide compared to traditional RS. Through a three-week diary study with 12 participants using custom GPTs for music recommendations, we found that LLM-powered CRS can (1) help users clarify implicit needs, (2) support unique exploration, and (3) facilitate a deeper understanding of musical preferences. Based on these findings, we discuss the new design space enabled by LLM-powered CRS and highlight its potential to support more personalized, user-driven recommendation experiences. }
\end{abstract}



\maketitle

\section{Introduction}

\textcolor{black}{A conversational recommender system (CRS) provides personalized recommendations through interactive conversations with users. Unlike traditional recommender systems (RS) that offer one-time suggestions, CRS enables users to discover desired recommendations through multi-turn interactions \cite{cai2022impacts, bursztyn2021developing, cai2020predicting}. The key value of CRS is that it transforms the user’s role from a passive recipient to an active participant in the recommendation process \cite{cai2022impacts, friedman2023leveraging, jannach2021survey, zhu2024reliable}. While this shift broadens the limited role of users in traditional RS, active participation does not necessarily guarantee user autonomy \cite{jin2019musicbot}. Existing CRS models are often constrained by pre-defined interaction patterns designed by developers or designers, and they tend to replicate the logic of training datasets \cite{zhang2024towards}, thereby failing to provide a recommendation service experience that fully reflects the diverse needs of users. }

\textcolor{black}{Against this backdrop, the development of Large Language Models (LLMs) has enabled open-ended interactions between users and the system, allowing for more flexible and unrestricted communication with the CRS \cite{zhao2023recommender}. This flexibility extends beyond improving system performance; it empowers users to shape their own modes of interaction, enabling them to actively design a personalized recommendation service experience. Specifically, users can: 1) explicitly specify the personal data they want the RS to consider, 2) clearly instruct the system on how to interpret this data and generate recommendations, and 3) determine how to provide feedback on the recommended items. This shift introduces a new dimension of personalized recommendation experiences distinct from those offered by traditional RS. However, most existing studies on LLM-powered CRS focus on leveraging LLMs to improve recommendation performance, with insufficient exploration of how users can create their own recommendation experiences or how these experiences differ from those of traditional RS. Accordingly, our study aims to address the following research questions: When users utilize an LLM-powered CRS to build their own personalized recommendation service, what unique quality of experience does this offer compared to traditional RS? }

To address these research questions, this study aimed to explore how users interact with a recommendation system in ways that align with their own preferences. To achieve this, a Customized CRS-GPT was developed for each participant, enabling them to freely interact with the system using their preferred interaction methods across three key stages of the recommendation process: preference elicitation, recommendation presentation, and user feedback. Participants engaged in a three-week diary study, during which they experienced music recommendations through their customized CRS-GPT. The findings revealed that this customized CRS allowed users to: (1) clarify implicit needs, (2) support unique exploration, and (3) facilitate a deeper understanding of musical preferences. Based on these findings, we discuss the new design space enabled by LLM-powered CRS.

\section{Related Work}

\subsection{Supporting active participation of users in CRS}

\textcolor{black}{In the field of Human-Computer Interaction (HCI), research has highlighted not only the positive aspects of RSs that offer everyday convenience but also the negative user experiences arising from the passive role imposed on users by RS \cite{eslami2019user, sullivan2019reading}. Traditional RS typically operates through a one-sided interaction structure where the system automatically analyzes user preferences and delivers recommendations without user intervention. This approach can narrow the user’s perspective, hinder exploratory behavior \cite{wilkinson2018testing, pariser2011filter, guesmi2022interactive, chu2020towards}, and lead users to passively accept the system's biased decisions \cite{wilkinson2018testing, lee2015personalization, knijnenburg2016recommender}. To overcome these limitations, CRSs have emerged, enabling users to explore recommended items through real-time, multi-turn conversations with the RS \cite{cai2022impacts, bursztyn2021developing, cai2020predicting}. Specifically, CRS allows users to explicitly request the types of recommendations they want \cite{cai2022impacts, friedman2023leveraging, jannach2021survey, zhu2024reliable} and provide feedback on recommended items over multi-turn interactions \cite{jin2019musicbot}. By offering users a channel to explicitly express their preferences, CRS introduces several benefits. Users’ explicit articulation of their preferences can help address the cold-start problem \cite{li2023exploring, zhai2024actions, dai2023uncovering, di2023evaluating, sanner2023large, jannach2021survey} and enable more personalized recommendations, thereby enhancing their tailored experience \cite{zhu2024reliable}. Additionally, this approach enhances the system's explainability, allowing users to better understand the rationale behind recommendation results \cite{li2023large, liu2023chatgpt, lubos2024llm, silva2024leveraging, gao2024generative}. Beyond these performance-oriented advantages, the introduction of CRS marks a significant shift in user experience, transforming users from passive recipients of recommendations into active participants in the recommendation process. }

\textcolor{black}{Despite these benefits, existing CRS often provided limited and fixed responses due to technical constraints \cite{zhang2024towards}, which restricted users from freely designing their desired recommendation service experience. For example, while users can offer feedback on items suggested by CRS, they are typically limited to predefined attributes such as energy, tempo, genre, or artist \cite{jin2019musicbot, cai2021critiquing}. Consequently, when users attempt to present more detailed or complex demands, the system may struggle to interpret or respond appropriately \cite{jin2019musicbot}. Against this backdrop, advancements in LLMs have opened the possibility of transcending the limited interaction methods of existing CRS. LLMs enable more flexible and open-ended interactions, allowing users to engage with CRS in a more dynamic and unrestricted manner, thereby introducing a new paradigm for recommendations \cite{sharma2024generative}. The following section provides a detailed discussion of these new possibilities. }

\subsection{LLM-powered CRS}

\textcolor{black}{The advancement of LLMs has significantly enhanced the natural language interface between users and systems, enabling machines to engage in human-like conversations \cite{friedman2023leveraging}. LLM-powered conversational services (e.g., ChatGPT, Gemini) can dynamically perform a wide range of functions in response to user prompts \cite{anelli2024sixth, petruzzelli2024towards, tankelevitch2024metacognitive, deng2023unified}, allowing users to create and utilize personalized interactive services tailored to their unique objectives \cite{kim2019co}. This development also presents new opportunities in the field of CRS \cite{petruzzelli2024towards}. Users are no longer confined to recommendation systems provided by specific companies; instead, they can design and operate their own customized recommendation services. For instance, the GPT store hosts a variety of user-developed recommendation systems with diverse objectives, enabling users to freely interact and collaborate with CRS. This approach allows users to design a personalized recommendation service experience according to their individual preferences \cite{petruzzelli2024towards, friedman2023leveraging}, providing a user experience distinct from traditional RS and CRS. This possibility transforms the three key interaction stages of a recommendation system \cite{pu2012evaluating, harambam2019designing}—1) preference analysis through user input, 2) recommendation presentation, and 3) user feedback—by giving users direct control over how they engage with the CRS at each stage. More specifically, LLMs expand the flexibility of the input space \cite{tankelevitch2024metacognitive}, allowing users to freely provide information that will be used in preference analysis. Users can also explicitly specify the types of recommendations they wish to receive \cite{li2023pbnr, weisz2024design} and determine how they want to evaluate the recommended items. This increased level of control enables users to further customize their recommendation experience, resulting in a new form of personalized recommendation experience that is clearly distinct from that of traditional RS. }

\textcolor{black}{On the other hand, prior studies have primarily focused on the potential of LLMs to improve recommendation performance, leaving a research gap in understanding this potential from the user's perspective. LLM-powered CRS enables users to fully determine the details of their interactions with the recommendation system, allowing them to go beyond simple explicit interaction. This approach empowers users to design and customize their own recommendation services. Given that individuals have diverse goals for using recommendation systems, providing users with the freedom to create personalized recommendation services opens up new possibilities for more tailored user experiences. This study views the potential of LLMs as an opportunity to explore the possibilities of customized CRS. Specifically, we aim to investigate how users want to interact with CRS, for what purposes, and what new user experiences such interactions can bring. }

\section{Method}

\subsection{Recommendation System Domain}

Before designing the study, we considered which recommendation domain would best help us achieve our research goals. Previous studies have explored user experiences with CRSs across various domains, including music, movies, books, news articles, and social media posts \cite{knijnenburg2016recommender, kunkel2019let, chu2020towards, harambam2019designing, sullivan2019reading, rader2018explanations, sinha2002role, sun2023recommender, di2023evaluating}. Among these diverse domains, we determined that the music recommendation domain was the most suitable for our study, based on two key factors. First, we aimed to observe active interactions between participants and the system, which required recommended items that could be quickly experienced and evaluated. Second, to allow for smooth comparisons with existing recommendation experiences, we needed to choose a domain that participants frequently engage with in their daily lives. Considering these factors, we concluded that the music domain was the most appropriate for achieving our research goals \cite{sinha2002role, sun2023recommender}. Therefore, in this study, we aim to use the music recommendation domain as a representative example for exploring LLM-powered CRS experiences.

\subsection{Participants}

We aimed to recruit participants who enjoy listening to music and frequently use music recommendation systems such as YouTube Music, Apple Music, or Spotify. Considering these criteria, along with gender and age balance, we ultimately recruited 12 participants (Table 1). \textcolor{black}{All recruited participants reported using music recommendation systems daily, with usage ranging from 30 minutes to as much as 4 hours per day. }Before participation, all recruits attended a pre-study session where they received detailed explanations about the study’s procedures and the types of data that would be collected. The study was conducted only with participants who provided their informed consent. We determined the compensation for this study based on the most common standards in the region where the study was conducted, and each participant was compensated with 150,000 KRW \textcolor{black}{(approximately 110 USD). }All aspects of this study were approved by the Institutional Review Board (IRB No.KH2023-262).

\begin{table*}[]
\caption{Recruited participants for the study}
\Description{include #, Gender, Age, List of music RS used, RS usage patterns}
\begin{tabular}{lllll}
\toprule[1pt]
\# & \textbf{Gender} & \textbf{Age} & \textbf{\begin{tabular}[c]{@{}l@{}}List of Music\\ RS used\end{tabular}} & \textbf{RS usage patterns} \\ \midrule[1pt]
P1 & Male & 27 & YouTube Music & \begin{tabular}[c]{@{}l@{}}Choose a song similar to one I want to hear from the recommended list,\\ then listen to the automatically played songs.\end{tabular} \\ \midrule[0.5pt]
P2 & Female & 24 & YouTube Music & \begin{tabular}[c]{@{}l@{}}Search for a song or select one from the recommended list,\\ then listen to the automatically played songs.\end{tabular} \\ \midrule[0.5pt]
P3 & Male & 27 & Spotify & \begin{tabular}[c]{@{}l@{}}Create playlists and use the smart shuffle feature to listen.\\ \end{tabular} \\ \midrule[0.5pt]
P4 & Female & 23 & \begin{tabular}[c]{@{}l@{}}YouTube Music,\\ Apple Music\end{tabular} & \begin{tabular}[c]{@{}l@{}}Occasionally use the recommendation feature,\\ but mostly listens to playlists\\ created by others or myself.\end{tabular} \\ \midrule[0.5pt]
P5 & Female & 26 & YouTube Music & \begin{tabular}[c]{@{}l@{}}Choose a song similar to one I want to hear from the recommended list,\\ then listen to the automatically played songs.\end{tabular} \\ \midrule[0.5pt]
P6 & Male & 28 & Spotify & \begin{tabular}[c]{@{}l@{}}Choose a song similar to one I want to hear from the recommended list,\\ then listen to the automatically played songs.\end{tabular} \\ \midrule[0.5pt]
P7 & Male & 23 & YouTube Music & \begin{tabular}[c]{@{}l@{}}Use the “Play again” feature to get recommendations similar to those frequently\\ played songs then listened using shuffle play.\end{tabular} \\ \midrule[0.5pt]
P8 & Male & 22 & YouTube Music & \begin{tabular}[c]{@{}l@{}}Use the shuffle feature to listen to songs similar to ones I've previously enjoyed.\\ \end{tabular} \\ \midrule[0.5pt]
P9 & Female & 25 & YouTube Music & \begin{tabular}[c]{@{}l@{}}Choose a song similar to one I want to hear from the recommended list\\ then listen to the automatically played songs.\end{tabular} \\ \midrule[0.5pt]
P10 & Female & 27 & \begin{tabular}[c]{@{}l@{}}YouTube Music,\\ Flo\end{tabular} & \begin{tabular}[c]{@{}l@{}}Listen mainly to newly released songs recommended by favorite artists or genres.\\ \end{tabular} \\ \midrule[0.5pt]
P11 & Female & 27 & YouTube Music & \begin{tabular}[c]{@{}l@{}}Listen to my own playlist, occasionally checking out recommended songs\\ on the home screen to add to my playlist.\\ \end{tabular} \\ \midrule[0.5pt]
P12 & Female & 28 & \begin{tabular}[c]{@{}l@{}}YouTube Music,\\ Flo, Spotify,\\ Apple Music\end{tabular} & \begin{tabular}[c]{@{}l@{}}Freely use various features, including YouTube Music's station feature\\ and Spotify podcasts.\end{tabular} \\ \bottomrule[1pt]
\end{tabular}
\end{table*}

\subsection{Diary Study Design}

In this section, we explain how we designed a three-week diary study to observe participants’ experiences with LLM-powered CRSs. The key challenge was determining how to effectively deliver the LLM-powered CRS experience while ensuring that participants could customize the recommendation process to suit their preferences and receive personalized recommendations. Given these considerations, we determined that custom GPTs were the most effective tool for our study, as they enabled participants to personalize their interactions with the system while also offering direct music recommendations.

To create custom GPTs for each participant, they first needed to decide how they wanted to interact with the system at each of the three stages of recommendation: preference analysis, recommendation presentation, and user feedback. \textcolor{black}{Since the concept of customizing the interaction with a CRS was relatively unfamiliar to participants, we chose not to introduce a finalized system. Instead, we aimed to design the study to allow them to explore diverse interaction possibilities, helping them familiarize themselves with the system’s open-ended capabilities while minimizing the influence of novelty or initial impressions. After sufficient exploration, we planned to have participants decide how they would prefer to interact with the recommendation system at each stage and to fully experience both the advantages and limitations of their chosen interaction methods. Furthermore, since these interaction methods could be continuously refined based on participants’ preferences, we intended to include an opportunity for them to revisit and adjust their initially chosen interaction style if they wished. This iterative process was designed to empower participants to refine their interactions and develop a deeper understanding of the system’s open-ended potential. }

To achieve these objectives, the study was structured into three distinct weeks. During Week 1, participants were introduced to various interaction possibilities across the three stages of recommendation (preferences analysis, recommendation presentation, and user feedback). This first week encouraged participants to explore and brainstorm different service scenarios through hands-on experience. In Week 2, participants selected their preferred recommendation service scenario from Week 1 and spent the week receiving recommendations based on that scenario. Finally, in Week 3, participants reflected on the strengths and weaknesses of their Week 2 scenario and were given the opportunity to modify it, if needed. They then spent the week experiencing recommendations based on the revised scenario. Through these three weeks, we aimed to help participants refine their needs while exploring the diverse possibilities of an LLM-powered CRS. The study, structured across three weeks, is shown in Figure 1. Further details of each week are explained in the following subsections.

\begin{figure}
    \centering
    \includegraphics[width=1\linewidth]{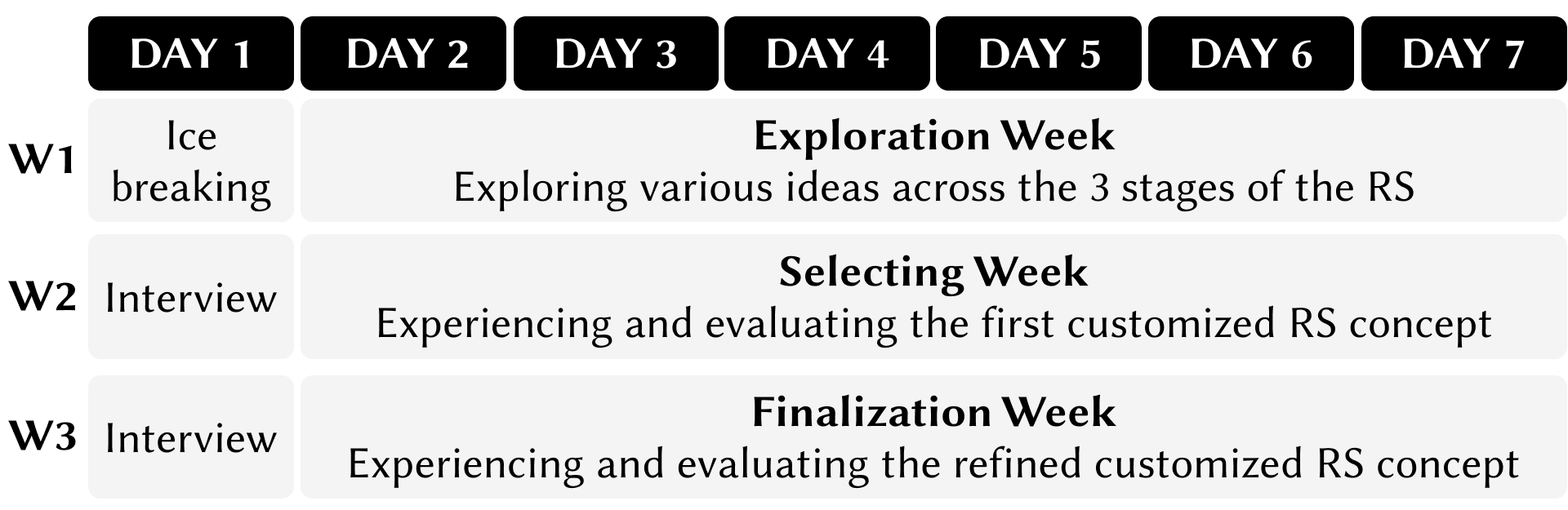}
    \caption{Three-week study schedule}
    \Description{On the first day of Week 1, there was an ice-breaking activity and Days 2 to 7 were for the Exploration Phase. In Week 2, the first day was for an interview, and Days 2 to 7 were for the Selecting Phase. In Week 3, the first day was for an interview, and Days 2 to 7 were for the Finalization Phase.}
    \label{fig:enter-label}
\end{figure}

\subsubsection{\textcolor{black}{Week 1: }Exploration \textcolor{black}{Week }– Exploring Various Ideas Across the Three Stages of Recommendation Process}

\textcolor{black}{Week 1 }was structured to provide participants with a comprehensive experience of different approaches across the three stages of the recommendation process. On the first day, participants engaged in an ice-breaking activity that allowed them to directly explore various interaction \textcolor{black}{scenarios }related to the three stages of RS. This activity was designed for participants to complete at their convenience from home. We provided them with a ChatGPT account and an ice-breaking sheet. Following the instructions on the sheet, participants interacted with ChatGPT to receive recommendations and then evaluated their experiences. Through these evaluations, we aimed to help participants brainstorm their own interaction \textcolor{black}{scenario }ideas.

In the \textbf{preferences analysis stage (Stage 1)}, participants were guided to explore different types of personal data that could be used for music recommendations. Drawing from prior research, we presented participants with the following four tasks:
\begin{itemize}
    \item Provide general demographic information about yourself (e.g., age, gender, occupation) and ask ChatGPT to recommend music. \cite{liao2022user, pazzani1999framework}
    \item List your five favorite songs and ask ChatGPT for music recommendations based on them. \cite{liao2022user, pazzani1999framework}
    \item Describe in detail what kind of music you like and ask ChatGPT for recommendations. \cite{kim2021teaching}
    \item Provide an image or photo that conveys a mood you want and ask for music recommendations that match the mood. \cite{park2024text}
\end{itemize}

Participants were asked to document the advantages and disadvantages of using each type of data on the icebreaking sheet. After completing these tasks, participants synthesized the advantages and disadvantages of the four data types and generated ideas about what kind of data would be most meaningful for making recommendations. Figure 2 shows an example of an icebreaking sheet completed by P5.

\begin{figure}%
\subfloat[]{{\includegraphics[width=0.47\textwidth ]{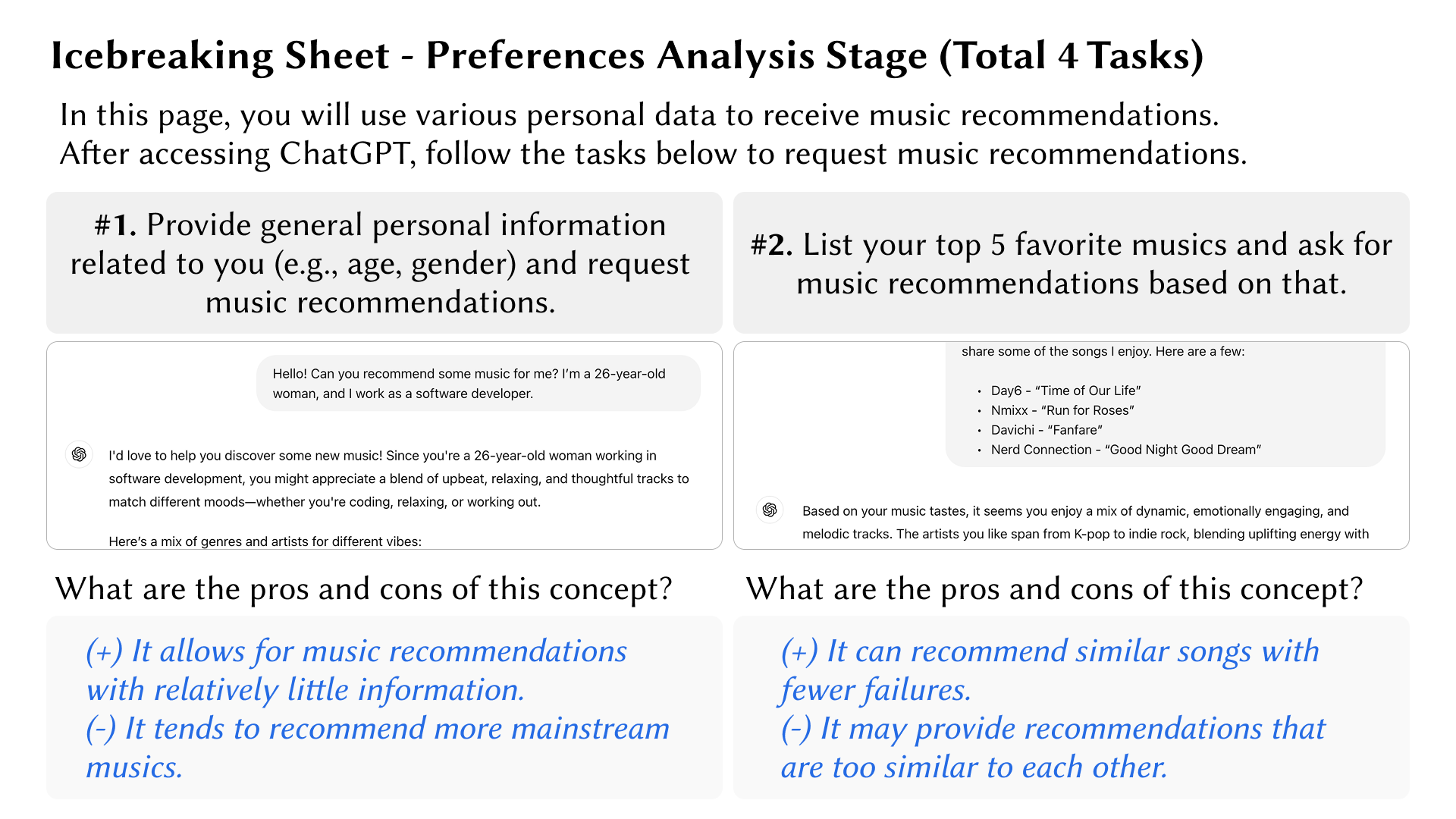} }}%
\hfill
\subfloat[]{{\includegraphics[width=0.47\textwidth ]{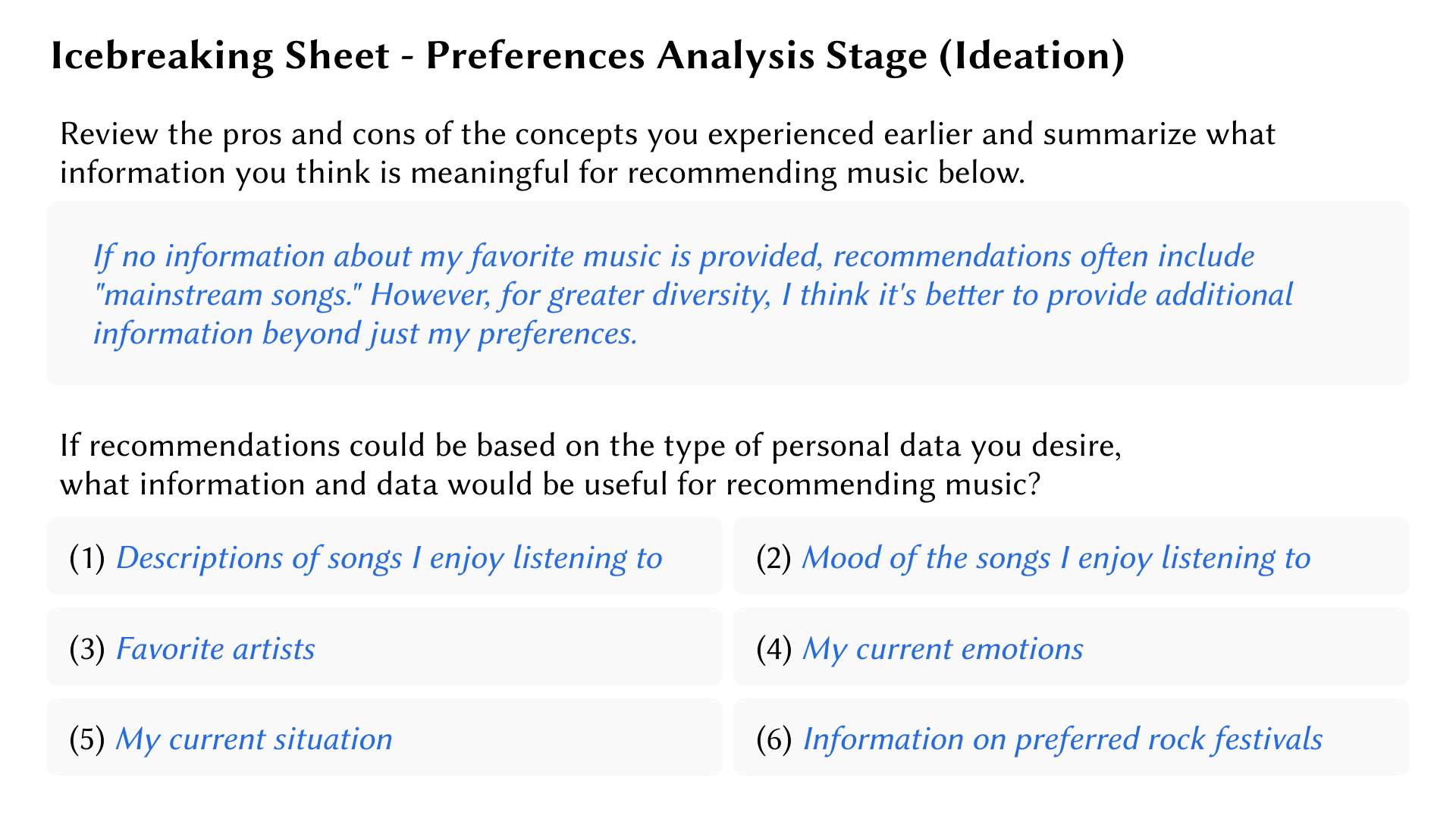} }}%
\caption{P5’s completed icebreaking sheet, (a) evaluating each scenario based on the researcher-provided tasks, (b) brainstorming personal ideas based on prior experiences}%
\Description{}
\label{tofsignal}%
\end{figure}

In the \textbf{recommendation presentation stage (Stage 2)}, participants were tasked with considering the purpose of the recommendations when the system selects items, as well as how they would like the reasons for those recommendations to be explained. After selecting one of the data types from Stage 1, participants were guided to provide that information to the recommendation system and complete the following three tasks. They then recorded the advantages and disadvantages of each \textcolor{black}{scenario }and brainstormed their own ideas on the icebreaking sheet:
\begin{itemize}
    \item Ask ChatGPT to recommend music you might dislike based on your information, and request a detailed explanation of why it was recommended. \cite{sullivan2019reading, knijnenburg2016recommender, balog2023measuring}
    \item Ask ChatGPT to recommend something you are likely unfamiliar with but might enjoy, with reasons provided using relevant keywords. \cite{sullivan2019reading, knijnenburg2016recommender, balog2023measuring}
    \item Ask ChatGPT to recommend something you might like, explaining the reasons with images. \cite{sullivan2019reading, knijnenburg2016recommender, jin2024exploring}
\end{itemize}

In the \textbf{user feedback stage (Stage 3)}, participants were introduced to various feedback methods. Using one data type from Stage 1, along with the recommendation objective and explanation styles chosen in Stage 2, they received recommendations and were asked to provide feedback in three different ways based on the tasks provided. They then recorded their thoughts on each feedback approach and brainstormed new ideas on the icebreaking sheet:
\begin{itemize}
    \item Rate the recommended music on a scale of 1 to 5. \cite{schnabel2020doesn}
    \item Request options from ChatGPT to explain why the recommendation is good or bad, and answer using a multiple-choice format. \cite{schnabel2020doesn}
    \item Provide a detailed explanation of why you liked or disliked the recommendation. Then, to encourage deeper reflection on your preferences, ask ChatGPT to generate additional questions. \cite{schnabel2020doesn}
\end{itemize}

After completing the first day’s activities, participants spent the next six days, from Day 2 to Day 7, receiving recommendations based on their own ideas they had generated during the ice-breaking activity. Each day, participants decided how they wanted to interact with the system in each of the three stages—(1) preference analysis, (2) recommendation presentation, and (3) user feedback—and completed a final prompt following the researcher’s guidance. Once the final prompt was complete, participants used ChatGPT to receive music recommendations. We asked participants to receive recommendations at least once a day, and after each activity, they were required to document two things in their Week 1 diary: (a) a screenshot of their interactions with the ChatGPT and (b) the advantages and disadvantages of the \textcolor{black}{scenario }they tested that day. Over the six days, participants were encouraged to fully experience all the ideas they generated on Day 1. Figure 3 shows an example of P6’s activity using ChatGPT to receive music recommendations.

\begin{figure}
    \centering
    \includegraphics[width=1\linewidth]{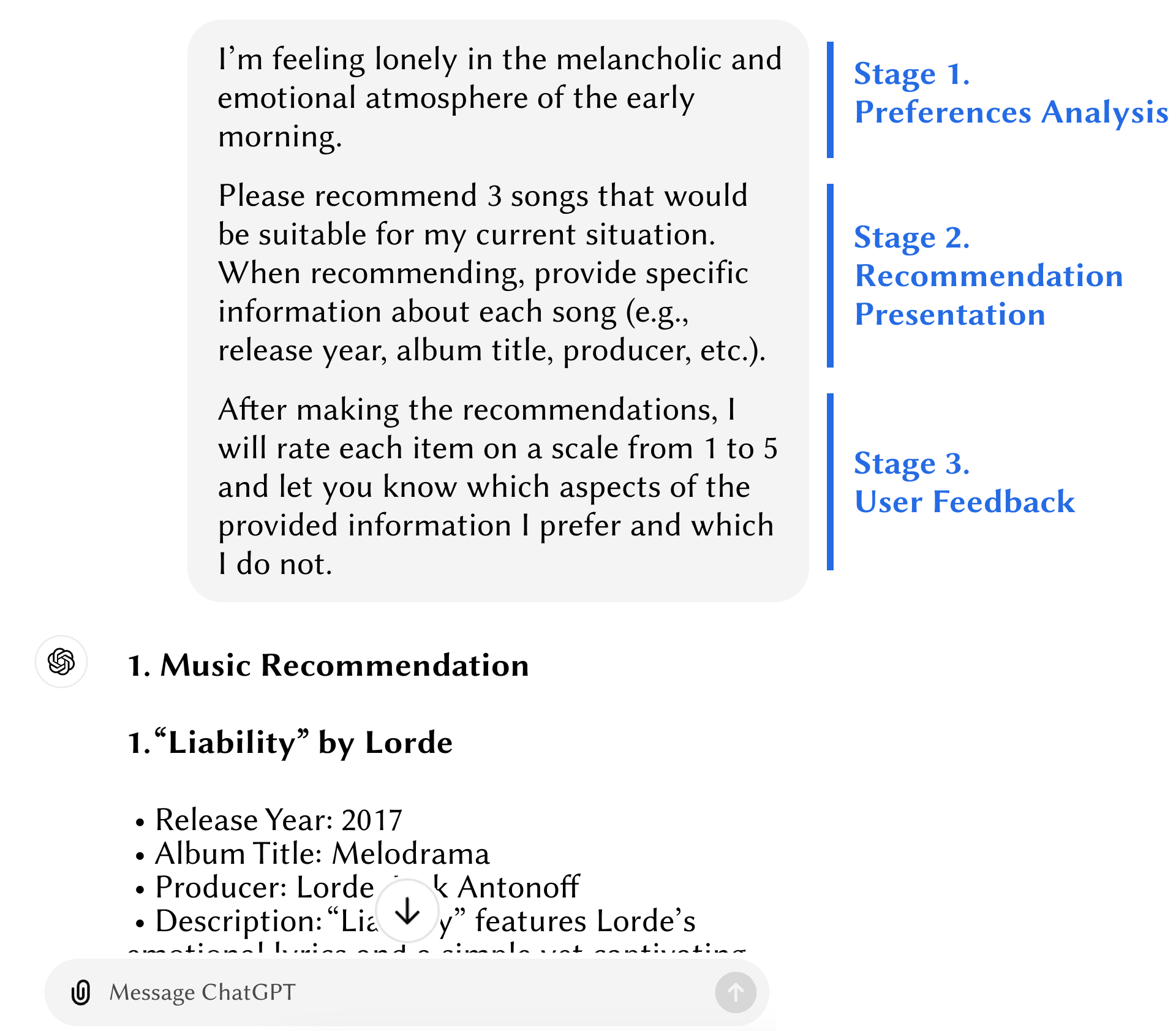}
    \caption{P6’s activity using ChatGPT for music recommendation}
    \Description{}
    \label{fig:enter-label}
\end{figure}

\subsubsection{\textcolor{black}{Week 2: }Selecting \textcolor{black}{Week }– Experiencing and Evaluating the First Customized Recommendation System}

In Week 2, participants consolidated the insights gained from Week 1 by synthesizing the advantages and disadvantages of various CRS scenarios they explored. From these, they selected the scenario they found most appealing and spent the second-week receiving recommendations based on their chosen scenario. To initiate this activity, the first day of Week 2 included a 20 to 30-minute interview with the researcher. The interview provided an opportunity for participants to reflect on their experiences from Week 1, focusing on the various ideas they tried across the three stages of interaction, the reasons for their preferences, and the interaction methods they found most meaningful. After the interview, participants finalized how they wanted to configure their interactions with the CRS for each stage and documented this on a planning sheet. The sheet included (1) the name of the CRS, (2) the objective of the chosen CRS service scenario, and (3) how they wanted to interact with the system in each of the three stages (preference analysis, recommendation presentation, and user feedback). Once participants had completed the planning sheet, the researcher used it to customize GPT. \textcolor{black}{When creating Custom GPTs, instructions were specified to ensure that GPT could function as a music recommendation system. Specifically, it was instructed to "analyze the user's music preferences based on the information provided by the user, recommend suitable music according to the analyzed preferences, and request feedback on the recommended music." }For example, P7 designed a system that recommended pop songs based on his preferences for K-pop and J-pop (Figure 4).

\begin{figure}
    \centering
    \includegraphics[width=1\linewidth]{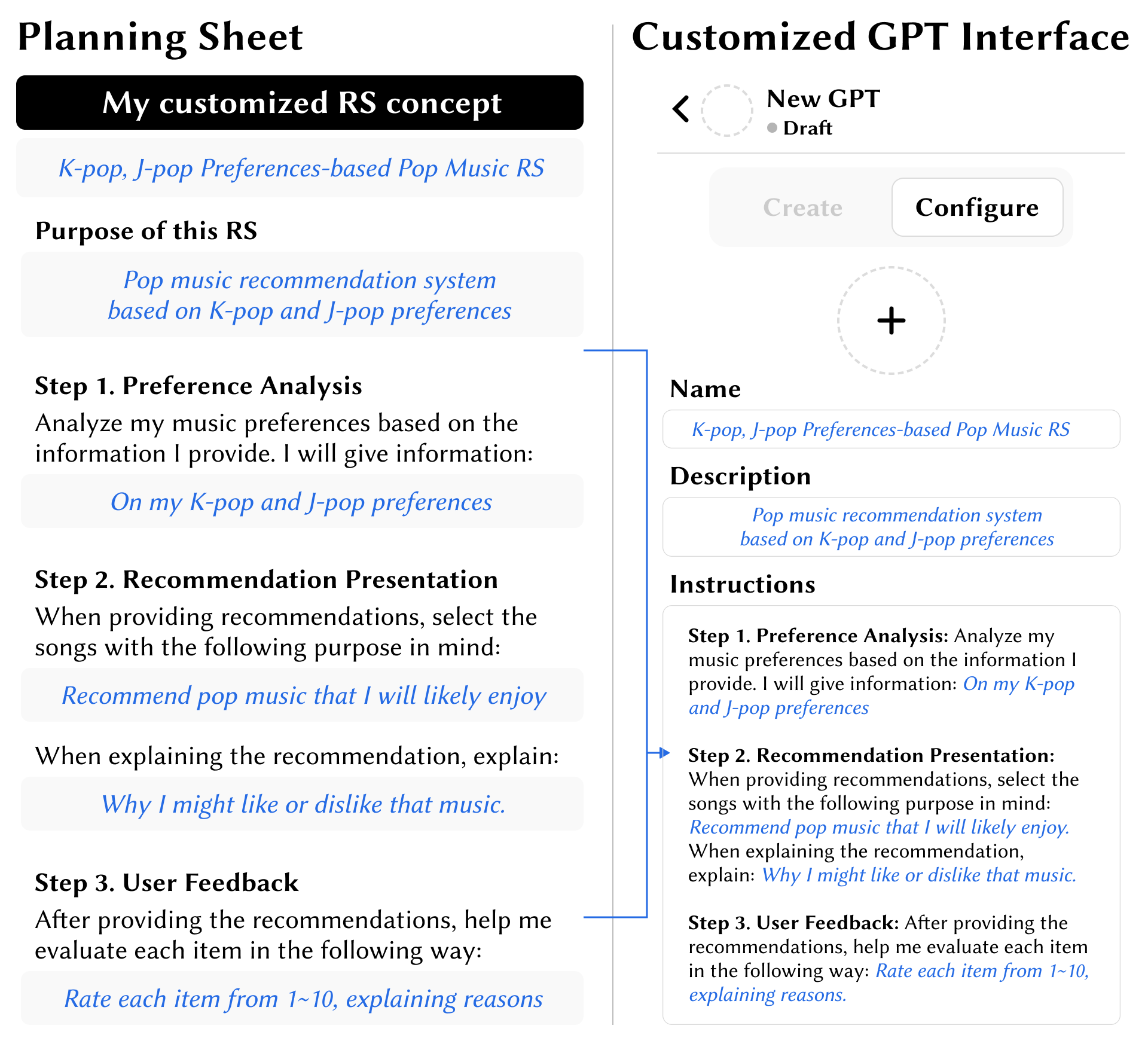}
    \caption{P7’s planning sheet and the GPT customization process (blue text indicates participant input)}
    \Description{P7 planned an RS called the “Pop music recommendation system based on K-pop and J-pop preferences.”}
    \label{fig:enter-label}
\end{figure}

During Days 2-7 of the second week, participants used their custom GPTs to receive music recommendations and recorded their experiences in the Week 2 diary. The diary included the following three components: (1) a screenshot of the interactions with custom GPT, (2) strengths and weaknesses of this scenario against their prior CRS experiences, and (3) suggestions for improvement. Figure 5 shows an example of P3 using the custom GPT to receive music recommendations.

\begin{figure}
    \centering
    \includegraphics[width=1\linewidth]{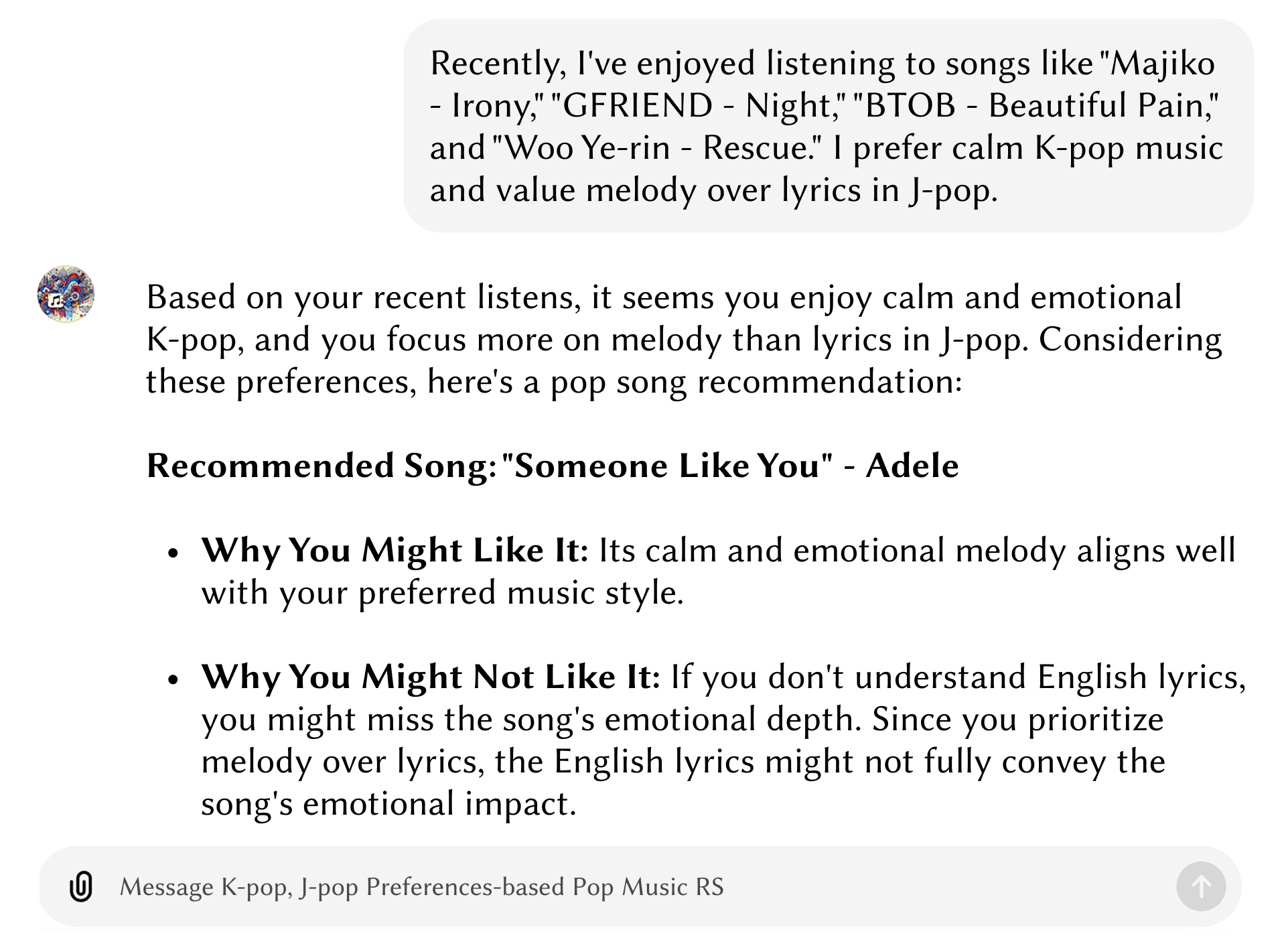}
    \caption{P3’s screenshot using the custom GPT to receive pop music recommendations based on preferences for K-pop and J-pop}
    \label{fig:enter-label}
\end{figure}

\subsubsection{\textcolor{black}{Week 3: }Finalization \textcolor{black}{Week }– Experiencing and Evaluating the Refined Customized Recommendation System}

The aim of Week 3 was to refine participants’ recommendation service scenarios based on their experiences from Week 2 and provide them with a week to interact with the improved scenario. On the first day of Week 3, participants attended a 20-30 minute interview with the researcher. During this interview, they reflected on their experiences with the LLM-powered CRS from the second week. Following the interview, participants received a planning sheet, identical in format to the one used in Week 2 (Figure 4). They used this sheet to outline how they would adjust their interactions with the system across the three stages (preference analysis, recommendation presentation, and user feedback). After submitting their updated plans, the researcher incorporated their feedback to refine and update their custom GPT models. From Days 2 to 7, participants used their updated custom GPTs to receive music recommendations and recorded their experiences in the Week 3 diary.

\subsection{Interview Procedure}

After the three-week study concluded, we conducted interviews with each participant to gather their experiences over the three-week period. The purpose of these interviews was to explore the potential advantages and challenges of LLM-powered CRS compared to traditional RS. The interviews were structured into three sections. In the first section, participants discussed how they typically use traditional music RSs, their favorite features, and the strengths and weaknesses of those systems. The second section focused on their experiences with the LLM-powered CRS during the study, drawing on their diary entries. In the final section, participants compared their experiences with the LLM-powered CRS to their previous experiences with traditional RS. Each interview lasted approximately 60 to 90 minutes.

\subsection{Overall Results and Data Analysis}

\textcolor{black}{This section presents the results obtained during our study. Section 3.5.1 describes the customized CRS scenarios created by the participants, while Section 3.5.2 explains the data analysis for qualitative findings from the participants' interview data. }

\subsubsection{Overall Results}

In Weeks 2 and 3, a total of 12 participants generated 16 distinct recommendation system scenarios tailored to various purposes. After completing Week 2, four participants opted to explore entirely new service scenarios for Week 3, creating two unique scenario ideas each. The remaining participants retained the overall direction and purpose of their original scenarios but adjusted specific interaction details for Week 3, resulting in the development of one refined scenario each. A summary of these scenario ideas and their purposes is presented in Table 2.

\begin{table*}[]
\caption{Recommendation system service scenarios developed for Weeks 2\&3 by 12 participants and purposes behind each \textcolor{black}{scenario }}
\Description{include Participants Number, Description of RS service scenario developed for Weeks 2 & 3, Purpose of the scenario}
\label{tab:my_table}
\begin{tabular}{lll}
\toprule[1pt]
\textbf{\#} & \textbf{\begin{tabular}[c]{@{}l@{}}Description of RS service scenario \end{tabular}} & \textbf{Purpose of the scenario} \\ \midrule[1pt]
P1 & \begin{tabular}[c]{@{}l@{}}Music RS based on favorite photos\\ \end{tabular} & \begin{tabular}[c]{@{}l@{}}To receive music recommendations even when it’s hard to describe\\desired music, by simply providing a favorite photo.\end{tabular} \\ \midrule[0.5pt]
P2& \begin{tabular}[c]{@{}l@{}}Music RS based on current situation\\ \end{tabular} & \begin{tabular}[c]{@{}l@{}}To explore new music that match the current situation.\\\end{tabular}\\ \cline{2-2}
 & \begin{tabular}[c]{@{}l@{}}Music RS that understands my current needs\\ \end{tabular} &  \\ \midrule[0.5pt]
P3 & \begin{tabular}[c]{@{}l@{}}Music RS based on my preferences\\ and personal information\end{tabular} & \begin{tabular}[c]{@{}l@{}}To broaden music preferences meaningfully by receiving recommendations \\ for new genres or songs that haven't been tried before.\end{tabular} \\ \midrule[0.5pt]
P4& \begin{tabular}[c]{@{}l@{}}Music RS based on virtual character information\\ \end{tabular} & \begin{tabular}[c]{@{}l@{}}To explore new music by using a virtual character's information, \\ diverging from personal tastes.\end{tabular} \\ \cline{2-3} 
 & \begin{tabular}[c]{@{}l@{}}Music RS based on disliked information\\ \end{tabular} & \begin{tabular}[c]{@{}l@{}}To prevent limited recommendations by only excluding\\ certain preferences, allowing more diverse music suggestions.\end{tabular} \\ \midrule[0.5pt]
P5 & \begin{tabular}[c]{@{}l@{}}Music RS based on rock festival information\end{tabular} & \begin{tabular}[c]{@{}l@{}}To prevent overly narrow recommendations by providing rock festival\\details rather than highly specific personal preferences.\\ \end{tabular} \\ \midrule[0.5pt]
P6& \begin{tabular}[c]{@{}l@{}}Dynamic music RS adjusting interaction\\ based on clarity of needs\end{tabular} & \begin{tabular}[c]{@{}l@{}}To receive flexible recommendations that adjust based on how clear\\  or unclear the user's music preferences are at any given moment.\\ \end{tabular} \\ \cline{2-3} 
 & \begin{tabular}[c]{@{}l@{}}Music RS expanding preferences\\ based on current situation\end{tabular} & \begin{tabular}[c]{@{}l@{}}To break away from fixed rules about what music\\ to listen to in certain situations by receiving\\ new music suggestions that fit the context.\end{tabular} \\ \midrule[0.5pt]
P7& \begin{tabular}[c]{@{}l@{}}Music RS based on comprehensive\\ personal information\end{tabular} & \begin{tabular}[c]{@{}l@{}}To explore new preferences by providing comprehensive\\ information and receiving broad recommendations.\end{tabular} \\ \cline{2-3} 
 & \begin{tabular}[c]{@{}l@{}}Music RS recommending pop songs\\ based on K-pop and J-pop preferences\end{tabular} & \begin{tabular}[c]{@{}l@{}}To discover new pop music preferences using\\ existing tastes in K-pop and J-pop as a foundation.\end{tabular} \\ \midrule[0.5pt]
P8 & \begin{tabular}[c]{@{}l@{}}Music RS that analyzes common\\ traits of favorite songs and\\ suggests similar music\end{tabular} & \begin{tabular}[c]{@{}l@{}}To better understand personal music preferences\\ and discover more songs that align with those traits.\end{tabular} \\ \midrule[0.5pt]
P9 & \begin{tabular}[c]{@{}l@{}}Music RS based on comprehensive\\ personal information\end{tabular} & \begin{tabular}[c]{@{}l@{}}To explore new preferences by providing broad,\\ inclusive personal information.\end{tabular} \\ \midrule[0.5pt]
P10 & \begin{tabular}[c]{@{}l@{}}Music RS expanding preferences\\ based on current situation and existing tastes\end{tabular} & \begin{tabular}[c]{@{}l@{}}To break personal listening habits and explore new music that fits\\ specific contexts and broadens tastes.\end{tabular} \\ \midrule[0.5pt]
P11 & \begin{tabular}[c]{@{}l@{}}Music RS that recommends based on\\personal preferences and refines those preferences\\using a tournament-style music evaluation\end{tabular} & \begin{tabular}[c]{@{}l@{}}To learn how to express musical preferences more accurately, \\ nabling better future recommendations.\end{tabular} \\ \midrule[0.5pt]
P12 & \begin{tabular}[c]{@{}l@{}}Music RS based on personality traits\end{tabular} & \begin{tabular}[c]{@{}l@{}}To expand music preferences by receiving\\ recommendations based on personality test results.\end{tabular} \\ \bottomrule[1pt]
\end{tabular}
\end{table*}

\subsubsection{Data Analysis}

From the study, we collected various forms of data, including the interaction scenario ideas that participants recorded at each stage in their icebreaking and planning sheets, as well as their diary entries, which documented their experiences with the LLM-powered CRS over the three-week period. These materials were used to help semi-structured the interview questions. \textcolor{black}{We aimed to conduct a qualitative analysis of user experiences, focusing on the participants' interview data. Since the first week was intended as a period for exploring various attempts, we excluded the interview data from the interview conducted on the first day of the second week (Section 3.3.2). Instead, we focused on the interview conducted on the first day of the third week (Section 3.3.3) and the final interview conducted after all activities were completed (Section 3.4) for our qualitative analysis. }The interviews were audio-recorded, and we collected approximately 912 minutes of interview data. All interviews were transcribed digitally, and key quotes were extracted. We then conducted a qualitative thematic analysis \cite{boyatzis1998transforming} of the interview data, and two HCI researchers with experience in qualitative data analysis performed open coding \cite{urquhart2013using}.

Initially, we established the following initial codes \textcolor{black}{(rightmost section of Figure 6)}: patterns of traditional RS usage, pros and cons of traditional RS, participants’ ideas for each of the three stages of LLM-powered CRS, the reason behind those ideas, and new possibilities and challenges introduced by LLM-powered CRS. After the first round of coding, we developed expanded codes to surface participants’ specific needs related to LLM-powered CRS. \textcolor{black}{In this study, participants were encouraged to create their own personalized recommendation service experiences. As a result, the specific interaction methods varied significantly across participants. Rather than focusing on the detailed differences in these methods, we aimed to code the fundamental goals that motivated participants to adopt such customized strategies. When these fundamental goals were related to addressing user experience issues they had encountered with traditional RS, we grouped the corresponding codes under a unified theme. }\textcolor{black}{Figure 6 presents an example of our coding procedures, illustrating the progression from initial codes to expanded codes, categorized final codes, and ultimately, the derived theme.} Based on this analysis, we identified three key opportunities of the LLM-powered CRS as discovered by the participants: (1) helping to clarify implicit needs, (2) \textcolor{black}{supporting unique exploration}, and (3) facilitating a deeper understanding of musical preferences. The next section provides a detailed explanation of these findings.

\begin{figure}
    \centering
    \includegraphics[width=1\linewidth]{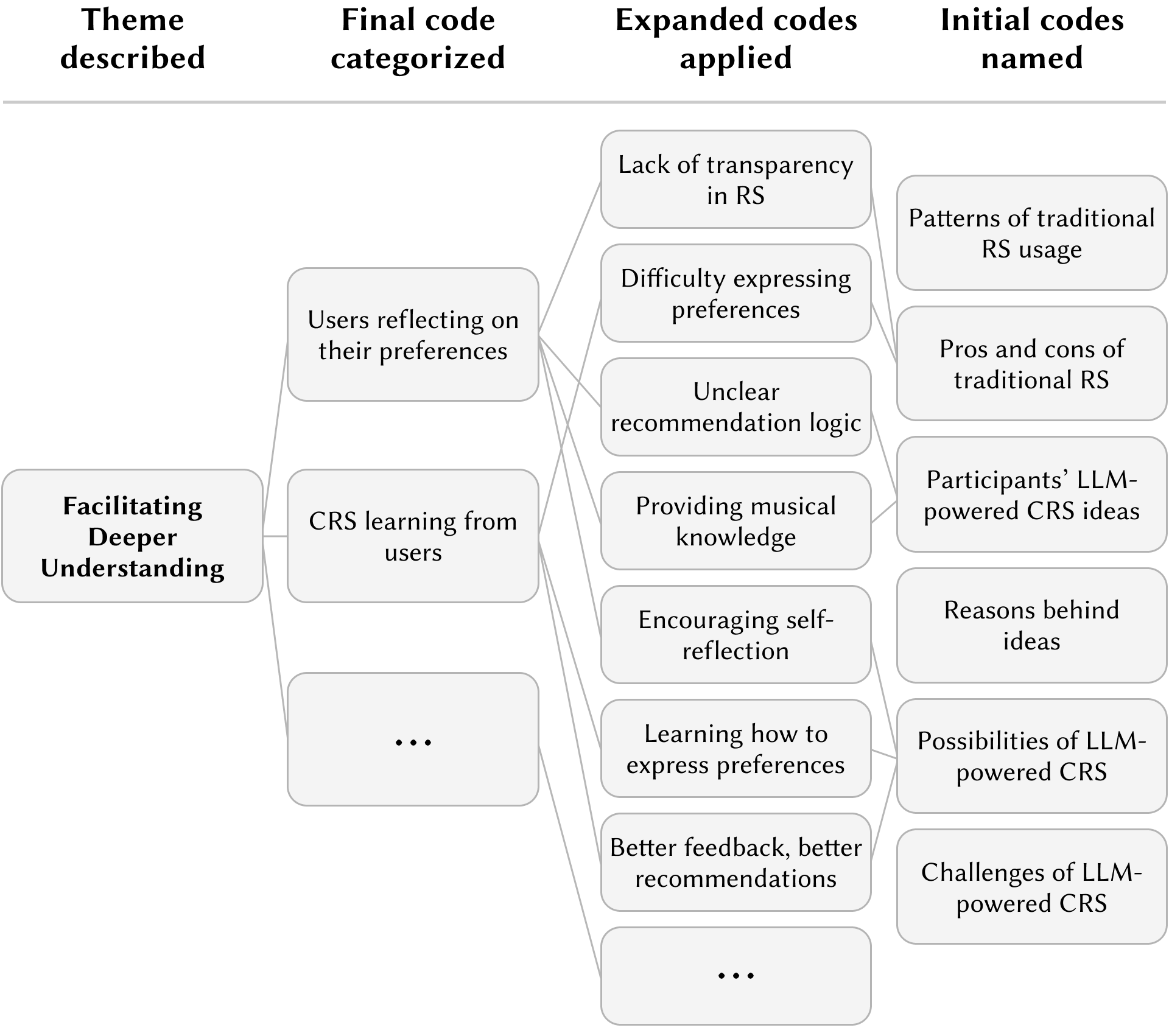}
    \caption{\textcolor{black}{Coding procedures for deriving the theme 'Facilitating Deeper Understanding' (selected examples)}}
    \label{fig:enter-label}
\end{figure}

\section{Findings}

\textcolor{black}{This section explains three key findings regarding the new quality of experience participants discovered in LLM-based CRS. Participants found that LLM-based CRS could provide novel recommendation experiences by helping to clarify implicit needs (Section 4.1), supporting unique exploration (Section 4.2), and facilitating deeper understanding (Section 4.3). In each sub-section, we describe the recommendation service scenarios in which participants experienced these possibilities, how these possibilities expanded traditional RS experiences, and what new concerns emerged.}

\subsection{Helping to Clarify Implicit Needs}

Participants could freely express their needs for music recommendations in their own language through a recommendation service tailored to their preferences. Participants often had implicit desires for the type of music they wanted to listen to, but they faced difficulties in clearly articulating these desires as specific musical preferences. To address this, they built a customized CRS to support the translation of their implicit needs, expressed in their own language, into concrete musical preferences.

When they had a desire to listen to a certain type of music but couldn’t clearly identify the specific song that matched it, they felt that traditional RSs did not provide sufficient support. This limitation led participants to settle for choosing a second-best option from the recommendations provided, rather than finding music that truly suited their needs.
\begin{quote}\textit{  
    When I don’t know exactly what I want to listen to, all I can do is click on a song from the lists. If I know what I want, I can search for it, but it’s when I have a vague desire that I think recommendations are most necessary. In those cases, though, the RSs don’t help much. (P10)}
\end{quote}

\textcolor{black}{The potential of LLMs offered participants a new user experience by enabling them to express their music preferences in diverse ways. For example, participants who found it difficult to describe the mood of the music they wanted to hear used images that conveyed a specific atmosphere, prompting the system to recommend music that matched the image. Similarly, participants who wished to discover music suited to their current situation but struggled to specify their musical preferences shared their thoughts, concerns, or emotions with the system. These examples highlight how LLM-powered CRS empowers participants to express their needs in various forms, allowing them to communicate naturally and intuitively. }
\begin{quote}\textit{ \textcolor{black}{
    It's hard for me to describe the mood I'm thinking of in detail. But I find it relatively easy to look for photos that capture the kind of mood I want. That's why I thought a system where I could show a picture that represents the mood of the music I want to listen to, and then get recommendations based on that, would be the most suitable for me. (P1)}}
\end{quote}
\begin{quote}\textit{  \textcolor{black}{
    I usually prefer finding and listening to music that fits my current situation. So, I came up with the idea of a service where I could explain my current situation to the recommendation system and receive recommendations based on that. Normally, if I identified my emotion, I would have to figure out which genre of music fits that mood and search for it myself. But with this system, I could skip that process and consider a broader range of options. (P6)}}
\end{quote}

Even when participants expressed their needs in their own way, the CRS effectively linked these inputs to music recommendations, providing meaningful suggestions. This process enabled participants to reflect on their true preferences through the recommended options, supporting them in uncovering their underlying desires. Rather than being a one-time experience, this exploration process offered participants an opportunity to learn which types of music would be effective in similar situations in the future. As a result, participants not only gained a deeper understanding of their own preferences but also discovered new possibilities for expanding their musical tastes.
\begin{quote}\textit{   
    Even when it’s just music for reading in a café, there are many different kinds of slower, softer music. Music with an ‘A’ vibe could work, and so could music with a ‘B’ vibe. By offering multiple playlists, it helped me think more about what I actually wanted. (P6)}
\end{quote}
\begin{quote}\textit{  \textcolor{black}{ 
    On this particular day, I received a really unique song recommendation, and it was an instrumental track with no lyrics. I had mentioned that I was feeling a bit stressed, and this song was recommended to me. As I listened to it, I found that it calmed me down much more than I had expected, so it really stuck with me. [...] It made me realize that when I want to feel this kind of emotion again, this type of song could be a good fit. (P2)}}
\end{quote}

However, while multi-turn interactions with the recommendation system were essential for gradually articulating participants' implicit needs, some participants found the extended recommendation process to be cumbersome. On the other hand, they expressed concerns that if the recommendation system were to automatically interpret and specify their implicit needs, it could increase their dependence on the system compared to traditional RS.
\begin{quote}\textit{  
    Compared to just clicking and listening, it can feel a bit inconvenient. (P4)}
\end{quote}
\begin{quote}\textit{   
    It was convenient that the system could give me satisfying recommendations even when I gave only a vague description of what I wanted. But I think I might become more dependent on it. Even if it suggested a genre I wasn’t originally interested in, I’d feel like ‘this is a recommendation made just for me,’ and that could make it harder to evaluate objectively. (P2)}
\end{quote}

In summary, the potential of CRS to allow participants to express their needs in their own language provided them with an experience that gradually facilitated the articulation of their implicit needs. However, it was found that careful attention must be paid to designing interactions with CRS that are not burdensome for users while also preventing excessive dependence on the system.

\subsection{Supporting Unique Exploration}

\textcolor{black}{The potential of open-ended interactions in LLM-powered CRS lies in the freedom to flexibly structure their interaction methods. By leveraging this capability, participants moved beyond the basic logic of traditional recommendation systems and developed their own unique recommendation logic. This enabled them to experience new music in a way that was more exploratory, meaningful, enjoyable, and accessible. }

Traditional RS, which recommends music similar to the songs participants frequently listen to, often limits the opportunity for participants to encounter new types of music that deviate from their expectations.
\begin{quote}\textit{ 
    The more I use the recommendation system, the more it feels like I’m just listening to familiar songs. Occasionally, a song I’ve never encountered before will appear on my recommendation list, but if it feels too random, I don’t feel motivated to give it a try. (P2)}
\end{quote}

\textcolor{black}{The LLM-powered CRS enabled participants to move beyond passively accepting system-generated recommendations by allowing them to define the purpose of the recommendations, specify the sources on which the system would base its suggestions, and determine how they intended to use the recommended music, thereby constructing diverse recommendation logic. For example, to explore preferences beyond their existing musical tastes, some participants aimed to reveal their musical preferences indirectly rather than stating them explicitly. P5 created a recommendation service that generated suggestions based on information about a specific rock festival she liked, while P12 built a service that utilized information from a music-related YouTube channel she frequently watched. Additionally, participants seeking to explore new types of music while maintaining accessibility built services that recommended songs from new genres or countries, using their existing preferences as a starting point. For example, P1 expanded his musical taste by receiving recommendations for Japanese rock ballads based on his well-established preference for K-pop rock ballads. Another participant, P3, developed a service that recommended songs beyond his existing preferences, introducing them to music from new eras, genres, and moods. }
\begin{quote}\textit{   
    I didn’t know much about J-pop, but I like rock ballads in K-pop, so I wanted recommendations for similar rock ballads in Japanese music. That’s why I asked for it. (P1) - Figure 7}
\end{quote}
\begin{quote}\textit{   
    When I asked for songs from different times, genres, or moods that matched my taste, the new recommendations I got felt fresh and still suited my preferences, which I really liked. (P3)}
\end{quote}

\begin{figure}
    \centering
    \includegraphics[width=1\linewidth]{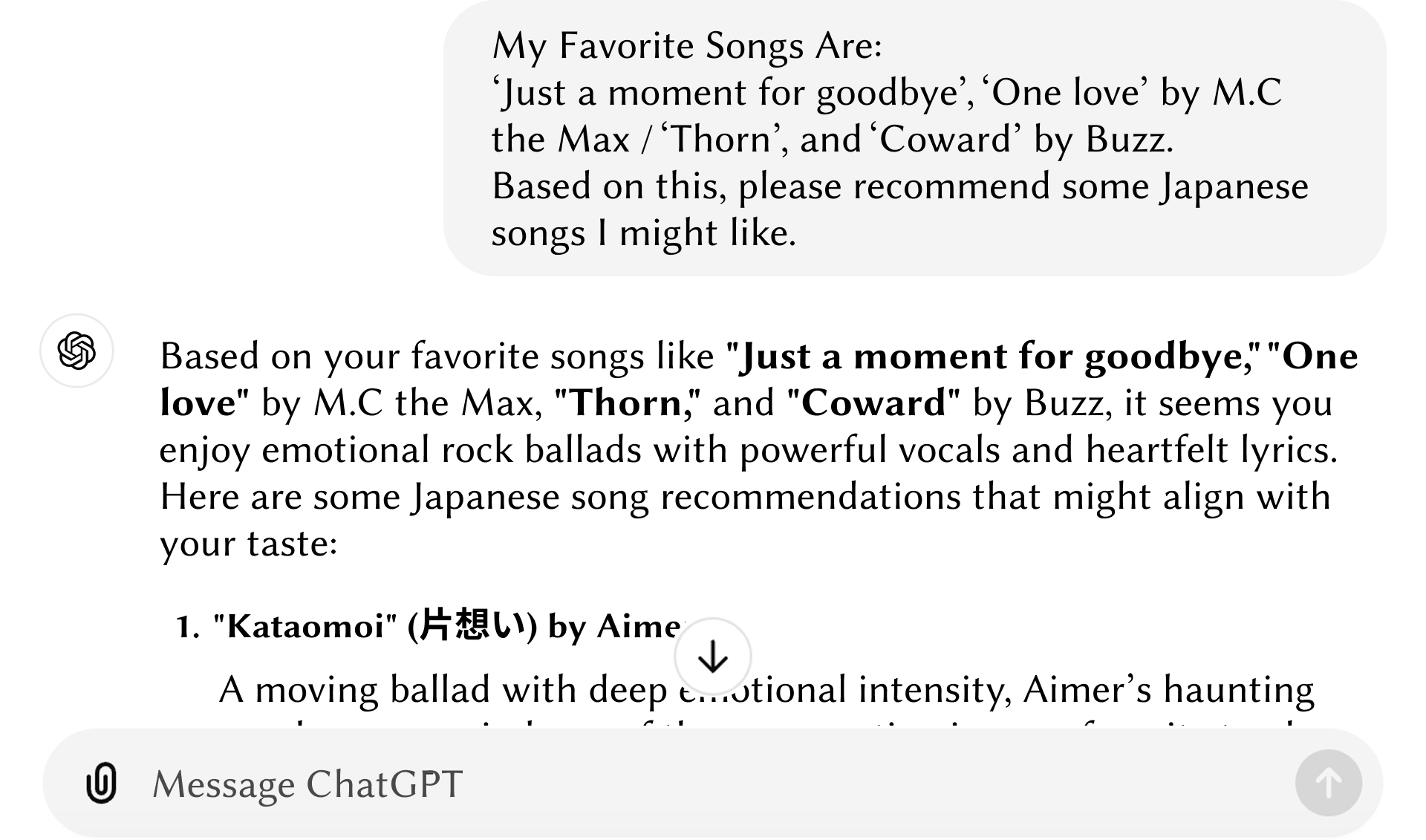}
    \caption{P1’s activity exploring similar J-pop based on his K-pop preferences}
    \label{fig:enter-label}
\end{figure}

\textcolor{black}{Furthermore, some participants designed the recommendation logic to suggest music not solely based on their musical preferences but on personal information such as their personality, hobbies, or daily routines. This approach enabled participants to enrich their daily experiences or use music as a tool for self-understanding. Rather than merely discovering new music, this process allowed participants to recognize new values in music, thereby expanding the role of traditional recommendation systems. }
\begin{quote}\textit{ \textcolor{black}{
    After I had just finished swimming in the morning, I wanted to keep that feeling going. So, I asked for music that could help me relive that moment and continue the lingering sense of it. (P2)}}
\end{quote}
\begin{quote}\textit{ \textcolor{black}{
    I received music recommendations based on my personality traits, and as I looked at the suggested songs, I started thinking, 'Why did the system think I would like this music?' That process made me feel a stronger attachment to the recommended music. (P9)}}
\end{quote}

\textcolor{black}{On the other hand, clearly explaining the rationale behind the recommendation process played a crucial role in helping participants understand the unique recommendation logic and motivating them to explore new music. However, when the provided rationale did not align with participants' expectations, it posed a risk of undermining trust in the recommendation system. For example, P5 received a song described as having a "refreshing vibe," but she felt that the song did not match this description, leading to a decline in her trust in the system. }

\textcolor{black}{In conclusion, the possibility of open-ended interactions with LLMs presents an opportunity to empower users to shape their own logic for personalized recommendation experiences. However, when the rationale behind the recommendations did not align with participants' expectations, it sometimes hindered their ability to derive meaningful value from the experience. }

\subsection{Facilitating Deeper Understanding}

Participants recognized the potential of LLM-powered CRS to go beyond simply providing recommendation items by offering supplementary information that supported them in discovering and articulating their own musical preferences. This information enabled participants to gain deeper insights into their tastes, fostering a collaborative experience where they could provide more specific and refined feedback to the system.

When using traditional RS, participants occasionally missed opportunities to reflect on their preferences, as they could not understand the reasoning behind certain music recommendations.
\begin{quote}\textit{   
    There were times when I wondered, 'Why did they recommend this song to me? What made them think I’d like this?' But I never got answers to those questions, so I didn’t really think deeply about it. (P4)}
\end{quote}

\textcolor{black}{Participants requested that the recommendation system provide information alongside music recommendations to help them clearly understand which elements influenced their preferences for certain songs. The type and depth of information requested varied among participants. Some participants sought objective musical characteristics, such as genre, era, or vocal style, to explore the underlying reasons for their preferences from a broader perspective. By using this information, they were able to objectively recognize and analyze their own musical tastes. In contrast, other participants preferred personalized explanations that highlighted the relationship between newly recommended music and their existing preferences, rather than focusing on objective attributes of the music itself. For example, participants requested insights into why they might like or dislike a newly recommended song based on their individual tastes. This approach allowed them to view the music from a more personal perspective and provided an opportunity to expand their existing preferences. }
\begin{quote}\textit{ \textcolor{black}{
    I can easily decide whether I like or dislike a song, but explaining why I like it is a completely different matter. The system explaining what genre the song is, what era it’s from, or what style it follows helped me think about what aspects I liked. It was helpful for organizing my preferences. (P12)}}
\end{quote}
\begin{quote}\textit{\textcolor{black}{
    When the system explained why I might like or dislike a particular song, I was able to view the new music in relation to my existing preferences. Even if I found myself liking aspects of the song that differed from the system’s explanation, it still felt meaningful, as it became an opportunity to update and expand my musical tastes. (P7)}}
\end{quote}

\textcolor{black}{Although the type and depth of information participants sought varied, the provision of such information enabled them to reflect on the factors that shaped their musical preferences. This highlights the potential of recommendation systems to evolve from mere recommendation tools into tools that support self-reflection and personal learning. Notably, this process marked a shift from a system-driven approach — where information is predefined and presented according to an algorithm — to a user-driven approach, where participants actively requested specific information they wanted to learn. }

Moreover, this process of better understanding their preferences also led to more precise and detailed feedback. Participants noted that as they gained a clearer understanding of their tastes, they were able to provide more valuable feedback, which in turn resulted in richer recommendations. This iterative process allowed both users and the CRS to evolve together, fostering a more refined and personalized recommendation experience.
\begin{quote}\textit{  
    When I understood what elements I liked, I immediately gave that information to the system, hoping for even better recommendations. (P6)}
\end{quote}
\begin{quote}\textit{  
    Being able to clearly explain my music preferences meant I could request music that matched my tastes even more accurately. Knowing my preferences was key to a better recommendation experience. (P11)}
\end{quote}

On the other hand, the system's overly persuasive analysis of participants' preferences sometimes led them to readily accept the suggested results as their own preferences. This, in turn, hindered participants from reflecting on and critically examining their own tastes.
\begin{quote}\textit{  
    When the recommendation system says, 'Based on my analysis, this is why you liked this song,' and explains it in such a convincing way, I can't help but think, 'Ah, so this must be my taste.' (P12)}
\end{quote}

In summary, LLM-powered CRS has the potential to support a collaborative learning experience where participants and the system jointly explore user preferences by providing personalized information that helps users organize their tastes. On the other hand, it was crucial to design the CRS's analysis in a way that encourages meaningful reflection, ensuring that participants do not passively accept the system's analysis without critical consideration.

\section{Discussion}

\textcolor{black}{In our study, we enabled open-ended interactions between users and CRS through the potential of LLMs, allowing users to create and utilize their own personalized recommendation services. This approach revealed that users could experience a user-driven form of RS, which supported clarifying implicit needs (Section 4.1), designing unique recommendation logic (Section 4.2), and facilitating a deeper understanding of preferences (Section 4.3). Based on these findings, we propose a new design space for human-centered CRS enabled by LLMs (Sections 5.1 and 5.2) and discuss the design considerations necessary to address potential issues that may arise from active user interaction (Section 5.3). }

\subsection{Support Self-Discovery with LLM-powered CRS}

\textcolor{black}{Traditional recommendation systems have efficiently delivered personalized content by collecting extensive preference information through user interactions. However, they have been less effective in supporting self-discovery — the process through which users reflect on, organize, and understand their own preferences. In contrast, our study demonstrates that LLM-powered CRS can overcome this limitation by serving as a tool for self-exploration and self-understanding. Our findings reveal that LLM-powered CRS enables users to articulate their current musical needs more concretely (Section 4.1) and supports them in organizing and refining their musical preferences (Section 4.3). Ultimately, this process facilitates users' self-discovery. Notably, these self-discovery experiences are well-aligned with the concepts of explorative search \cite{white2009exploratory, marchionini2006exploratory} and sense-making processes \cite{russell1993cost, pirolli2005sensemaking, qu2008model}, where users actively explore, interpret, and integrate new information to gain a deeper understanding of their preferences. }

\textcolor{black}{LLM-powered CRS provided users with the opportunity to clarify their musical needs, even in situations where they were unable to explicitly articulate their preferences. By presenting a variety of options that users might like, the system enabled them to better understand and specify what they were looking for. This experience aligns with the goals and context of exploratory search, which emphasizes open-ended exploration where users evaluate and synthesize diverse information sources to achieve a deeper understanding of a topic \cite{white2009exploratory, marchionini2006exploratory}. LLM-powered CRS supported users' exploratory search by understanding and responding to vague or undefined needs unrelated to specific musical preferences (Section 4.1). This approach aligns with previous research efforts that aimed to support users' exploratory search using LLMs, where systems propose higher- or lower-level topics that users might find beneficial to explore \cite{fok2024marco, palagi2017survey, qu2008model}. Through this exploratory process, users were able to discover new insights about their musical preferences, such as identifying which types of music might be suitable for similar future situations. This process was particularly meaningful, as it allowed users to engage in self-discovery of their musical tastes. Previous studies have also explored methods to help users structure and organize such insights into a more formalized form, enabling them to refer back to this knowledge for future searches \cite{chang2019searchlens, crescenzi2021supporting}. Similarly, LLM-powered CRS could provide users with a space to structure and articulate their newly discovered internal knowledge, offering a way to support users' future exploratory searches. }

\textcolor{black}{Furthermore, LLM-powered CRS played a crucial role in helping users integrate newly discovered preferences into their existing preference system (Section 4.3). By providing analyses of specific musical elements in a song based on users' preference information, LLM-powered CRS enabled users to recognize which aspects of a song they particularly liked. This process supported users' sense-making — the ability to connect newly acquired information with their existing knowledge system and identify patterns and relationships between information \cite{russell1993cost, pirolli2005sensemaking, qu2008model}. This approach aligns with previous studies that aimed to support users' sense-making by analyzing relationships between pieces of information \cite{suh2023sensecape, jiayu2024jamplate, kang2023synergi, zheng2024disciplink, lee2024paperweaver}. It also highlights how LLM-powered CRS enables users to move beyond passively receiving information to actively reconstructing and expanding their self-understanding. Additionally, sense-making in the context of musical preferences differs from simple knowledge exploration, as it involves internal, subjective knowledge of personal taste. To support this process, it may be useful to incorporate diverse perspectives, such as drawing on the tastes of close friends \cite{kwak2024investigating}, perspectives from people with entirely different preferences \cite{bhuiyan2022othertube}, or reflections on one’s own past preferences. These strategies could promote a more diverse and holistic understanding of musical preferences, fostering deeper self-reflection. }

\textcolor{black}{The ability of LLM-powered CRS to facilitate exploratory search activity—where users explore the content space in diverse ways—and sense-making activity—where they deepen and refine their understanding of their own preferences based on newly acquired insights—suggests that LLM-powered CRS can serve a purpose beyond simply helping users find the desired content. This highlights the potential for LLM-powered CRS to function as a self-discovery supporter. Accordingly, this perspective suggests that designers should view LLM-powered CRS as a tool that supports users in exploring and understanding their own preferences. By adopting this approach, designers can develop recommendation systems that actively facilitate users' self-discovery, shifting the role of RS from a content delivery system to a system that promotes self-understanding. }

\subsection{Support User’s Designability in LLM-powered CRS}

\textcolor{black}{As described in Section 4.2, providing users with designability—the ability to create their own customized recommendation service experience—enabled them to shape unique recommendation logic that aligned with their personal values and needs (Section 4.2). This approach allowed users to define the recommendation system in a way that held deeper meaning for them, rather than merely following system-defined logic. This designability offered users diverse recommendation experiences. For instance, users could control how specifically they revealed their musical preferences to the system. By intentionally providing only indirect information about their preferences, users could actively adjust the accuracy of the recommendations. Additionally, users explored recommendations using personal data beyond musical preferences, enabling them to discover music with greater personal value, rather than simply consuming music suggested for general enjoyment. These experiences went beyond the goal of achieving "accurate recommendations" alone. Even when the recommended music did not align with their existing preferences, users were able to engage in reflective experiences, where the new music prompted them to reconsider their daily lives or rediscover aspects of their personal identity. This shift highlights the potential for LLM-powered CRS to support not only content discovery but also self-reflection. }

\textcolor{black}{LLM-powered CRS supports users in defining the purpose of recommendations, selecting the data sources to achieve that purpose, and deciding how to utilize the recommendation results. Through this process, users move beyond the role of passive recipients of recommendations. Instead, they actively design and construct their own recommendation logic, thereby defining the value of the recommendation system according to their personal goals and needs. Notably, users take on the role of service designers, crafting a personalized recommendation experience tailored to their unique preferences. This aspect of designability in LLM-powered CRS allows users to flexibly alter the intended goals of the system and directly determine its ultimate purpose. While traditional recommendation systems are often influenced by the interests of other stakeholders, such as corporations or other users, LLM-powered CRS enables users to construct their own algorithms free from external influence. This autonomy is particularly significant, as users can design their recommendation experience according to their own logic, reflecting their personal values and preferences. For example, users can exclude factors they do not wish to be influenced by (e.g., marketing-driven content or artists involved in social controversies). They can also incorporate their own important values — values often overlooked in commercialized recommendation systems — such as well-being, safety, or long-term goals \cite{stray2024building}. This approach allows users to create more meaningful and personalized recommendation experiences, where the recommendation logic is fully aligned with their individual values and goals. }

\textcolor{black}{This potential aligns with prior research, which emphasizes the importance of allowing users to define and control the purpose of the system \cite{kim2023investigating}. LLM-powered CRS serves as a flexible design tool that embraces this possibility. Building on this potential, recommendation system designers should move away from approaches that impose rigid interaction strategies on users. Instead, they should explore new design strategies that empower users to recognize their own preferences and needs and actively design and control the system based on this understanding. This shift can foster a more user-driven recommendation system experience, where users transition from passive recipients to active creators of their CRS experience. By enabling users to define the meaning and purpose of the CRS on their own terms, designers can support the development of personalized and value-driven CRS experiences. This approach allows users to create a recommendation experience that reflects their individual values, goals, and unique logic, ultimately leading to a more meaningful and personalized engagement with the system. }

\subsection{Support Opportunities for User Interpretation Through Ambiguity}

\textcolor{black}{As discussed in Sections 5.1 and 5.2, explicit and open interaction between the recommendation system and users has the potential to provide a new, user-driven recommendation experience. However, several issues were also identified that could negatively impact the user experience. For example, while LLM-powered CRS enables explicit communication with users to propose recommendations, an overly logical and persuasive recommendation process might discourage users from exploring and clarifying their own ambiguous needs (Section 4.1). Additionally, there is a risk that users may passively accept the system's analysis of their preferences without engaging in reflection (Section 4.3). Conversely, if the system’s recommendations lack sufficient empathy or fail to provide convincing reasoning, users may lose trust in the system, highlighting the dual nature of these interactions (Section 4.2). These challenges could limit the potential of LLM-powered CRS to support active user participation. Therefore, a key consideration for future system design is how to leverage explainability in the recommendation process to adjust the level of persuasion and user reflection \cite{radensky2023think}. Exploring strategies that encourage users to reflect on their own preferences while ensuring trust in the recommendation system will be an important direction for future research.}

\textcolor{black}{Previous studies have explored strategies to encourage users to move beyond passively accepting AI system explanations and instead critically evaluate AI reasoning through analytical thinking \cite{wang2021explanations, zhang2022towards, schoeffer2024explanations, buccinca2021trust, bansal2021does, lee2023understanding}. Among these, Sivertsen et al. \cite{sivertsen2024machine} proposed a strategy for conversational system design that emphasizes intentional ambiguity rather than striving for complete transparency in AI explanations. This approach aims to promote user autonomy and critical thinking by encouraging users to interpret the meaning of ambiguous and uncertain explanations on their own. This perspective is particularly relevant to LLM-powered CRS, which, despite having the capability to fully reveal the recommendation process, can instead leave room for users to explore and interpret the meaning of their recommendation experience. By doing so, LLM-powered CRS can foster a user-driven recommendation experience, where users actively engage in the process of meaning-making. At the same time, CRS can adopt a flexible interaction model to accommodate users with varying needs. For users who prefer to actively explore and reflect on their preferences, the system can leverage ambiguity to encourage exploration and critical thinking. Conversely, for users who prefer a more straightforward experience, CRS can reduce ambiguity and provide more specific, concrete information about the recommendation process. By balancing these interaction strategies, LLM-powered CRS can adapt to diverse user needs and offer a more personalized and meaningful recommendation experience.
In conclusion, we propose that, for the design of user-driven recommendation experiences, intentionally incorporating ambiguous interactions in CRS can positively impact the user experience by leaving room for users to generate their own interpretations. This approach could encourage users to actively engage in meaning-making, fostering a more reflective and personalized recommendation experience. }

\section{Limitations \& Future Work}

The custom GPTs used in this study were highly efficient tools, as they allowed us to easily customize the interaction methods according to each participant’s preferences. However, several limitations were identified during the study. First, since the custom GPTs were not integrated with real music applications, they sometimes recommended songs that didn’t exist or provided inaccurate information due to a lack of detailed music data. Another limitation lies in the participant recruitment process. \textcolor{black}{In order to select participants with both active experience using RSs and a basic understanding of GPT, the study primarily involved participants in their 20s, which limited the diversity of the sample. }Lastly, the study was conducted over a relatively short period of three weeks. If the study were extended over a longer period to observe how people build new preferences from start to finish using LLM-powered RSs, it could provide valuable insights into how usage patterns of these systems gradually evolve. \textcolor{black}{Furthermore, in designing the Method section, we aimed to mitigate the novelty factor's influence by allowing participants sufficient time to explore and familiarize themselves with the system. However, a three-week period was not long enough to completely eliminate the novelty effect. Therefore, conducting a longer-term study would be valuable from this perspective as well. Given that users’ music preferences are not established in a short period, our findings suggest that the process of self-discovery regarding one’s own preferences may be better observed in a long-term study. Additionally, it would be interesting to investigate how users modify the CRS they initially designed over time. Observing these modifications could offer valuable insights into the iterative nature of user-system interactions in the context of LLM-powered CRS. }

Building on these limitations, future research could focus on designing interactions that support the three opportunities of LLM-powered RS in greater detail. Developing systems that can be integrated with actual music applications would enable long-term studies with a more diverse group of participants in everyday settings.

\section{Conclusion}

In this study, we explored user experiences with LLM-powered RSs that enable active interaction between users and the system, examining the new opportunities and challenges these systems present. Over three weeks, we conducted a diary study in which participants used custom GPTs to interact with the recommendation system in ways they preferred, and documented their music recommendation experiences. Our findings revealed that participants expected three key opportunities from LLM-powered CRS: (1) helping to clarify implicit needs, \textcolor{black}{(2) supporting unique exploration}, and (3) facilitating a deeper understanding of musical preferences. Based on these findings, we discussed how these opportunities can redefine the design space and expand user roles in RS. Through this study, we hope to contribute to future research that emphasizes the importance of user-centered roles in recommendation systems and encourages mutually beneficial interactions between users and these systems.

\begin{acks}
We thank all participants of our user studies, as well as Hankyung Kim and Minseo Park for their valuable support and feedback. This work was supported by the National Research Foundation of Korea(NRF) grant funded by the Korea government(MSIT) (No. NRF-2021R1A2C2004263).
\end{acks}

\bibliographystyle{ACM-Reference-Format}
\bibliography{sample-base}

\end{document}